\documentclass[%
 preprint,
superscriptaddress,
notitlepage,
showpacs,preprintnumbers,
 amsmath,amssymb,
 aps,
floatfix,
]{revtex4-1}

\usepackage[T1]{fontenc}
\usepackage[english]{babel}
\usepackage[parfill]{parskip}

\usepackage{textgreek}
\usepackage{color}
\usepackage[caption=false]{subfig}

\usepackage{amsmath}
\usepackage{graphicx}
\usepackage{booktabs}
\usepackage{tabularx}
\usepackage{natbib}
\usepackage{dcolumn}
\usepackage{bm}
\usepackage{xspace} 

\usepackage[colorinlistoftodos]{todonotes}

\newcommand{\nova}{NO{\textnu}A}
\newcommand{\minerva}{MINER{\textnu}A}
\newcommand{\gevc}{\mbox{GeV/$c$}\xspace}

\newcommand{\eV}{\ensuremath{\mbox{e\kern-0.1em V}}\xspace}
\newcommand{\GeV}{\ensuremath{\mbox{Ge\kern-0.1em V}}\xspace}
\newcommand{\MeV}{\ensuremath{\mbox{Me\kern-0.1em V}}\xspace}
\newcommand{\GeVc}{\ensuremath{\mbox{Ge\kern-0.1em V}\!/\!c}\xspace}
\newcommand{\GeVcc}{\ensuremath{\mbox{Ge\kern-0.1em V}\!/\!c^2}\xspace}
\newcommand{\AGeV}{\ensuremath{A\,\mbox{Ge\kern-0.1em V}}\xspace}
\newcommand{\AGeVc}{\ensuremath{A\,\mbox{Ge\kern-0.1em V}\!/\!c}\xspace}
\newcommand{\MeVc}{\ensuremath{\mbox{Me\kern-0.1em V}/c}\xspace}

\newcommand{\dd}{\ensuremath{{\mathrm{d}}}\xspace}
\newcommand{\dedx}{\ensuremath{\dd E\!/\!\dd x}\xspace}

\newcommand{\ePl}{\ensuremath{e^+}\xspace}

\newcommand{\pip}{\ensuremath{\pi^+}\xspace}

\newcommand{\mup}{\ensuremath{\mu^+}\xspace}

\newcommand{\hp}{\ensuremath{\textup{h}^+}\xspace}

\newcommand{\CernVM}{\textsc{Cern\-\kern-0.05emVM}\xspace}

\begin{document}


\title{Measurements of total production cross sections for \boldmath{$\pi^{+}$}+C, \boldmath{$\pi^{+}$}+Al, \boldmath{$K^{+}$}+C, and \boldmath{$K^{+}$}+Al at 60\,\gevc and \boldmath{$\pi^{+}$}+C and \boldmath{$\pi^{+}$}+Al at 31\,\gevc}


\affiliation{National Nuclear Research Center, Baku, Azerbaijan}
\affiliation{Faculty of Physics, University of Sofia, Sofia, Bulgaria}
\affiliation{Ru{\dj}er Bo\v{s}kovi\'c Institute, Zagreb, Croatia}
\affiliation{LPNHE, University of Paris VI and VII, Paris, France}
\affiliation{Karlsruhe Institute of Technology, Karlsruhe, Germany}
\affiliation{Fachhochschule Frankfurt, Frankfurt, Germany}
\affiliation{University of Frankfurt, Frankfurt, Germany}
\affiliation{Wigner Research Centre for Physics of the Hungarian Academy of Sciences, Budapest, Hungary}
\affiliation{University of Bergen, Bergen, Norway}
\affiliation{Jan Kochanowski University in Kielce, Poland}
\affiliation{Institute of Nuclear Physics, Polish Academy of Sciences, Cracow, Poland}
\affiliation{National Centre for Nuclear Research, Warsaw, Poland}
\affiliation{Jagiellonian University, Cracow, Poland}
\affiliation{AGH - University of Science and Technology, Cracow, Poland}
\affiliation{University of Silesia, Katowice, Poland}
\affiliation{University of Warsaw, Warsaw, Poland}
\affiliation{University of Wroc{\l}aw,  Wroc{\l}aw, Poland}
\affiliation{Warsaw University of Technology, Warsaw, Poland}
\affiliation{Institute for Nuclear Research, Moscow, Russia}
\affiliation{Joint Institute for Nuclear Research, Dubna, Russia}
\affiliation{National Research Nuclear University (Moscow Engineering Physics Institute), Moscow, Russia}
\affiliation{St. Petersburg State University, St. Petersburg, Russia}
\affiliation{University of Belgrade, Belgrade, Serbia}
\affiliation{University of Geneva, Geneva, Switzerland}
\affiliation{Fermilab, Batavia, USA}
\affiliation{Los Alamos National Laboratory, Los Alamos, USA}
\affiliation{University of Colorado, Boulder, USA}
\affiliation{University of Pittsburgh, Pittsburgh, USA}


\author{A.~\surname{Aduszkiewicz}}
\affiliation{University of Warsaw, Warsaw, Poland}
\author{E.V.~\surname{Andronov}}
\affiliation{St. Petersburg State University, St. Petersburg, Russia}
\author{T.~\surname{Anti\'ci\'c}}
\affiliation{Ru{\dj}er Bo\v{s}kovi\'c Institute, Zagreb, Croatia}
\author{B.~\surname{Baatar}}
\affiliation{Joint Institute for Nuclear Research, Dubna, Russia}
\author{M.~\surname{Baszczyk}}
\affiliation{AGH - University of Science and Technology, Cracow, Poland}
\author{S.~\surname{Bhosale}}
\affiliation{Institute of Nuclear Physics, Polish Academy of Sciences, Cracow, Poland}
\author{A.~\surname{Blondel}}
\affiliation{University of Geneva, Geneva, Switzerland}
\author{M.~\surname{Bogomilov}}
\affiliation{Faculty of Physics, University of Sofia, Sofia, Bulgaria}
\author{A.~\surname{Brandin}}
\affiliation{National Research Nuclear University (Moscow Engineering Physics Institute), Moscow, Russia}
\author{A.~\surname{Bravar}}
\affiliation{University of Geneva, Geneva, Switzerland}
% ...

\collaboration{NA61/SHINE Collaboration}
\noaffiliation

\date{\today}  

\begin{abstract}

This paper presents several measurements of total production cross sections and total inelastic cross sections for the following reactions: $\pi^{+}$+C, $\pi^{+}$+Al, $K^{+}$+C, $K^{+}$+Al at 60\,\gevc, $\pi^{+}$+C and $\pi^{+}$+Al at 31\,\gevc .    The measurements were made using the NA61/SHINE spectrometer at the CERN SPS.  Comparisons with previous measurements are given and good agreement is seen.  These interaction cross sections measurements are a key ingredient for neutrino flux prediction from the reinteractions of secondary hadrons in current and future accelerator-based long-baseline neutrino experiments.

\end{abstract}


\maketitle


\section{Introduction}

The NA61 or SPS Heavy Ion and Neutrino Experiment (SHINE)~\cite{na61detector} at the CERN Super Proton Synchrotron (SPS) has a broad physics program that includes heavy ion physics, cosmic ray physics, and neutrino physics.  Long-baseline neutrino beams are typically initiated by high-energy protons that strike a long target, yielding hadrons that can decay to neutrinos or can reinteract in the target or in the aluminum focussing horns, potentially producing additional neutrino-yielding hadrons.  NA61/SHINE has already been very successful at measuring the yields of secondary hadrons generated by 31\,\gevc protons on carbon targets~\cite{na61_t2k_thin, na61_t2k_long} for the Tokai-to-Kamioka (T2K) long-baseline neutrino oscillation experiment~\cite{t2knim}.   Data at higher energies are now being collected to benefit other neutrino experiments, particularly \minerva~\cite{minerva_nim}, \nova{}~\cite{nova_design} that use the current NuMI neutrino beamline at Fermilab, and the proposed DUNE experiment~\cite{dune_physics} which will use the planned LBNF beamline.  The NuMI beamline is initiated by 120\,\gevc  protons on a carbon target, while LBNF will use 60-120\,\gevc protons on a carbon or beryllium target.  

In addition to the interactions of the primary protons in the neutrino beam targets, a significant fraction of the neutrinos result from hadrons coming from the reinteractions of 10-60\,\gevc energy pions, protons, and kaons in the carbon target, aluminum horns, and other beamline materials.    For example, in the on-axis low-energy beam configuration in NuMI, there are on average  $\sim$1.4 hadronic interactions per $\nu_{\mu}$~\cite{numi_flux}.  For the current medium-energy NuMI beam configuration and for DUNE, this increases slightly.  Uncertainties on the total cross sections for these ancestor hadrons results in uncertainties on the total hadron production rate and production location.  Especially for kaon interactions, the existing data is somewhat limited and not very well-reproduced by Monte Carlo~\cite{numi_flux}.

During the fall of 2015, NA61/SHINE recorded interactions of positively charged protons, pions, and kaons on thin carbon and aluminum targets. In the case of pions, interactions were recorded at beam momenta of 31\,\gevc and 60\,\gevc. Kaons were recorded with a beam momentum of 60\,\gevc only, and protons at 31\,\gevc only.  The NA61/SHINE vertex magnets were not operational during this period. Therefore, final state particles could not be identified and spectral measurements could not be extracted from this data run. As a result of this setup,  data-taking was optimized for making measurements of the total production and total inelastic cross sections for each interaction.  In the future, NA61 will extract hadron production spectra from data collected more recently with magnetic fields.

The total cross section of hadron-nucleus interactions $\sigma_\mathrm{tot}$ can be defined in terms of the inelastic $\sigma_\mathrm{inel}$ and coherent elastic $\sigma_\mathrm{el}$ cross sections:
\begin{eqnarray}
\sigma_\mathrm{tot} = \sigma_\mathrm{inel} + \sigma_\mathrm{el}.\label{eq:tot_xsec}
\end{eqnarray} 
The inelastic cross section $\sigma_\mathrm{inel}$ is defined as the sum
of all processes due to strong interactions except coherent nuclear elastic scattering.   The production processes are defined as those in which new hadrons are produced. The inelastic processes additionally include interactions which
only result in the disintegration of the target nucleus (quasi-elastic
interactions). Taking into account quasi-elastic scattering as a subset of the inelastic scattering process, one can define the production cross section $\sigma_\mathrm{prod}$ in terms of the quasi-elastic cross section $\sigma_\mathrm{qe}$ as:
\begin{eqnarray}
\sigma_\mathrm{prod} = \sigma_\mathrm{inel} - \sigma_\mathrm{qe}.\label{eq:prod_xsec}
\end{eqnarray} 

This paper is organized as follows:
Section~\ref{sec:Setup} describes the experimental apparatus.
Section~\ref{sec:evt} presents the event selection to ensure the quality of the measurements.  
Section~\ref{sec:ine_xsec} presents 
the procedure for measuring $\sigma_\mathrm{inel}$ and $\sigma_\mathrm{prod}$ cross sections.
Section~\ref{sec:corr_factors} describes the corrections to the raw trigger probability.
Section~\ref{sec:systematics} discusses systematic uncertainties. 
The final results and discussion are presented in Sections~\ref{sec:results} and~\ref{sec:Discussion}.

\section{Experimental setup, Beams, and Data Collected}\label{sec:Setup}

\begin{figure*}[t]
  \centering
  \includegraphics[width=\textwidth]{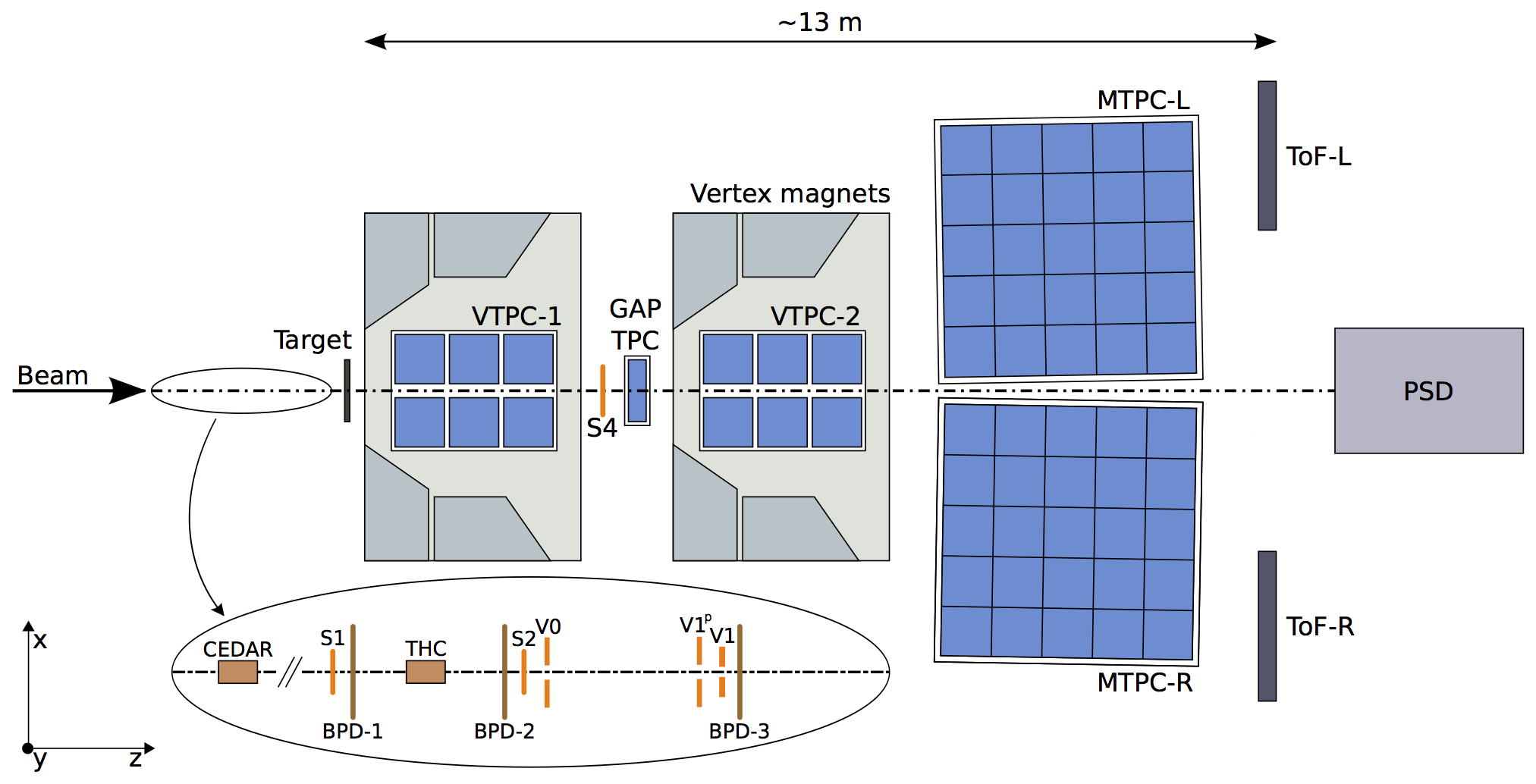}
 \caption{The schematic top-view layout of the NA61/SHINE experiment in the configuration used during the 2015 data-taking.  The two superconducting vertex magnets were not operational during the data-taking period.}\label{fig:exp_setup}
\end{figure*}

 NA61/SHINE receives a secondary hadron beam from the 400\,\gevc SPS proton beam. The primary proton beam strikes a beryllium target 535\,m upstream generating the secondary beam. A magnet system is then used to select the desired beam momentum. Unwanted positrons and electrons are absorbed by two 4\,mm lead absorbers.  
  
The NA61/SHINE detector~\cite{na61detector} is shown in Figure~\ref{fig:exp_setup}.   In standard operation, it comprises four large Time Projection Chambers (TPCs)  and a Time of Flight (ToF) system allowing NA61/SHINE to make spectral measurements of produced hadrons. Two of the TPCs, Vertex TPC 1 (VTPC-1) and Vertex TPC 2 (VTPC-2), are contained within superconducting magnets, capable of generating a combined maximum bending power of 9 T$\cdot$m. However these magnets were not operational during the 2015 run presented here.  Downstream of the VTPCs are the Main TPC Left (MTPC-L) and Main TPC Right (MTPC-R).  Additionally, a smaller TPC, the Gap TPC (GTPC), is positioned along the beam axis between the two VTPCs.   Two side time-of-flight walls, ToF-Left and ToF-Right, walls were present. The Projectile Spectator Detector (PSD), a forward hadron calorimeter, sits downstream of the ToF system.  

\begin{figure*}[tb]
\begin{center}
\includegraphics[width=0.48\textwidth]{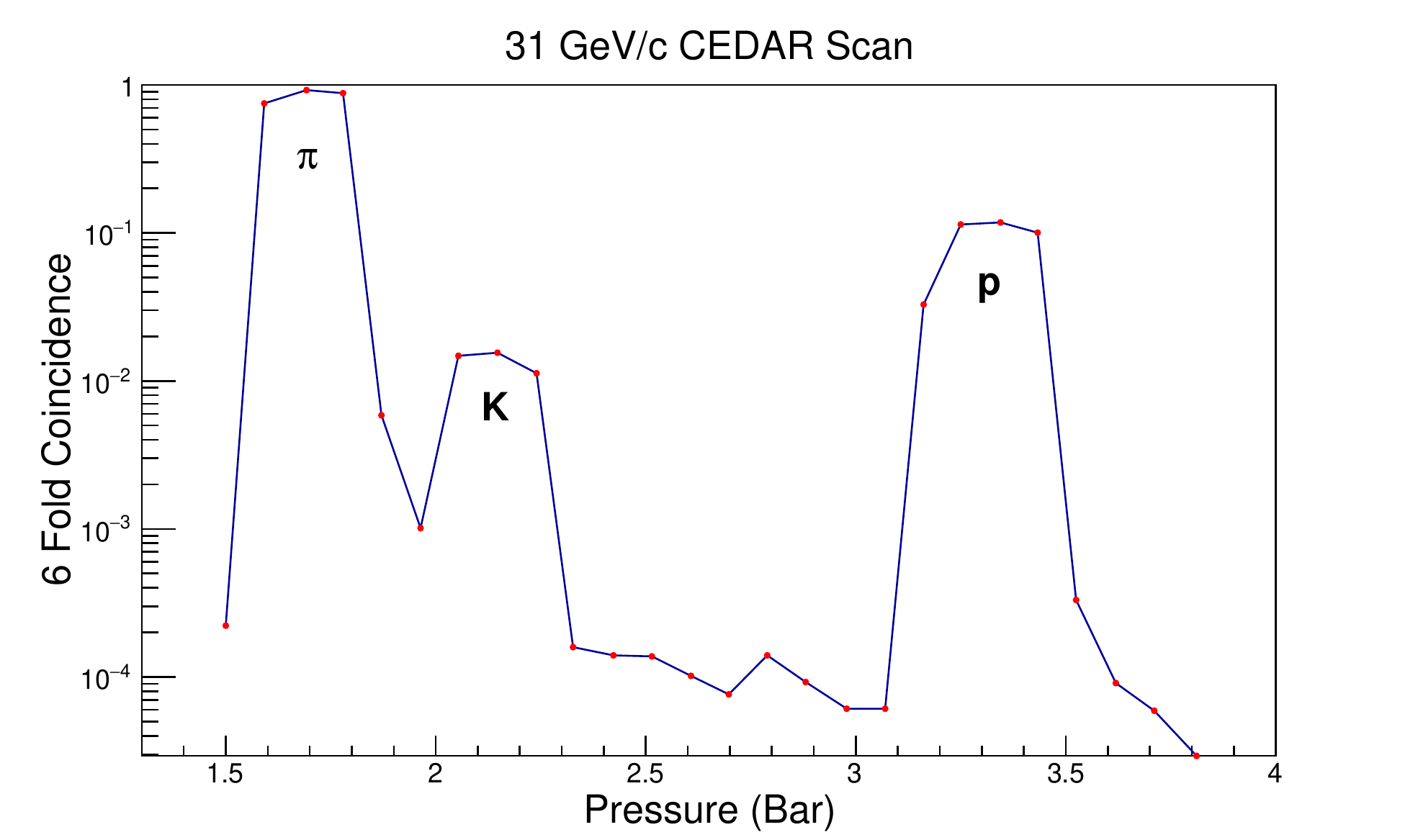}
\includegraphics[width=0.48\textwidth]{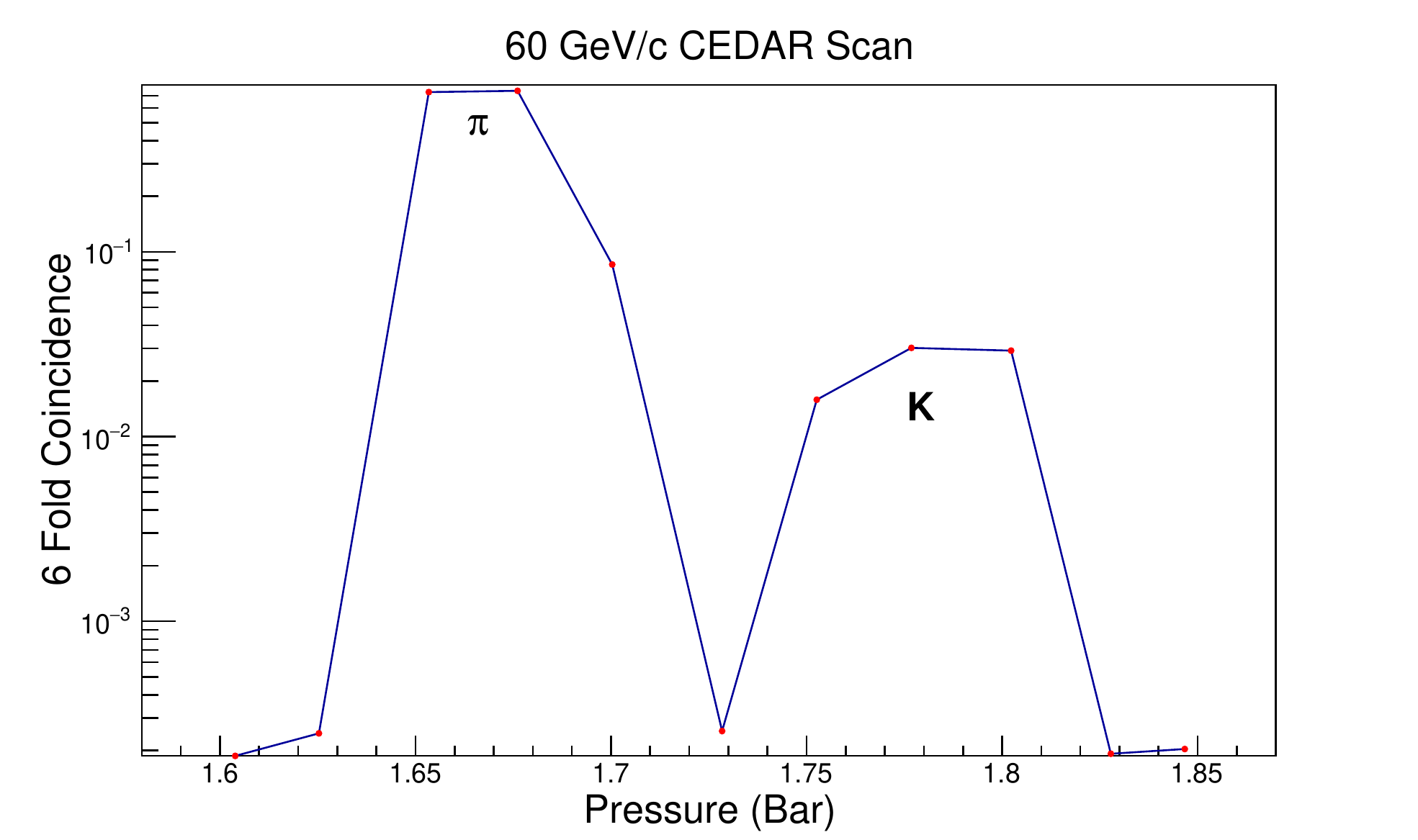}
\caption{CEDAR pressure scans for the 31 \gevc beam ($left$) and the 60 \gevc beam ($right$).  The vertical axis shows the fraction of beam particles that fires at least 6 of the 8 CEDAR PMTs. The green dashed lines show the approximate values of the CEDAR pressure settings that were used for the data sets analyzed here.}
\end{center}
\label{fig:cedar_scans}
\end{figure*}

The most critical systems for the analyses of the 2015 data presented here are the trigger system and the Beam Position Detectors (BPDs).   The NA61/SHINE trigger system uses two scintillator counters (S1 and S2) to trigger on beam particles. The S1 counter provides the start time for all counters. Three veto scintillation counters ($V0$, $V1$ and $V1^{p}$) each with a 1 cm diameter hole are used to remove divergent beam particles upstream of the target. The S4 scintillator with a 1 cm radius sits downstream of the target and is used to determine whether or not an interaction has occurred.   A  Cherenkov Differential Counter with Achromatic Ring Focus (CEDAR)~\cite{cedar, cedar82} and a threshold Cherenkov counter (THC) select beam particles of the desired species.  The CEDAR focusses the Cherenkov ring from a beam particle onto a ring of 8 PMTs. The pressure is set to a fixed value so that only particles of the desired species will trigger the PMTs, and typically a coincidence of at least 6 PMTs is required to tag a particle for the trigger.   Pressure scans taken of the CEDARs are shown in Figure~\ref{fig:cedar_scans}.  For these 2015 data at 31 \gevc the beam was composed of approximately 87\% pions, 11\% protons, and 2\% kaons, and the CEDAR pressure was set to 1.7 bar for the pion beam data, and 3.32 bar for the proton beam data.  At 60 \gevc the beam was composed of approximately 74\% pions, 23\% protons, and 3\% kaons. The CEDAR pressure was set to 1.78 bar for the kaon beam data and 1.68 bar for the pion beam data.

The beam particles are selected by defining the beam trigger ($T_\mathrm{beam}$) as the coincidence of $S1\wedge S2\wedge\overline{V0}\wedge\overline{V1}\wedge\overline{V1^{p}}\wedge CEDAR\wedge\overline{THC}$.
The interaction trigger ($T_\mathrm{int}$) is defined by the coincidence of $T_\mathrm{beam}\wedge\overline{S4}$ to select beam particles which have interacted with the target.  A correction factor will be discussed in detail in Section~\ref{sec:s4_xsec} to correct for interactions that hit the S4.  Three BPDs, which are  proportional wire chambers, are located 30.39 m, 9.09 m, and  0.89 m upstream of the target and determine the location of the incident beam particle to an accuracy of $\sim$100\,$\mu$m.

For these 2015 data, the interactions of $p$, $\pi^{+}$, and $K^{+}$ beams were measured on thin carbon and aluminum targets.  The carbon target was composed of graphite of density $\rho = 1.84\, \mbox{g/cm}^{3}$ with dimensions of 25\, mm (W) x 25\, mm (H) x 20\, mm (L), corresponding to roughly 4\% of a proton-nuclear interaction length.  The aluminum target has a density of  $\rho = 2.70\, \mbox{g/cm}^{3}$ with dimensions of 25\, mm (W) x 25\, mm (H) x 14.8\, mm (L), corresponding to roughly 3.6\% of a proton-nuclear interaction length.

\section{Analysis Procedure}\label{sec:evt}
\subsection{Event selection}

Several cuts were applied to events to ensure the purity of the measurement and to control the systematic effects caused by beam divergence. First, the so-called WFA (Wave Form Analyzer) cut was used to remove events in which multiple beam particles pass through the beam line in a small time frame. The WFA determines the timing of beam particles that pass through the S1 scintillator, with a resolution of 100 nsec. If another beam particle passes through the beam line close in time to the triggered beam particle, it could cause a false trigger in the S4 scintillator. In order to mitigate this effect, a conservative cut of $\pm$ 2 $\mu$s was applied to the time window to ensure that only one particle is allowed to pass through the S1 in a 4 $\mu$s time window around the selected beam particle. 

\begin{figure*}[tb]
\begin{center}
\includegraphics[width=0.45\textwidth]{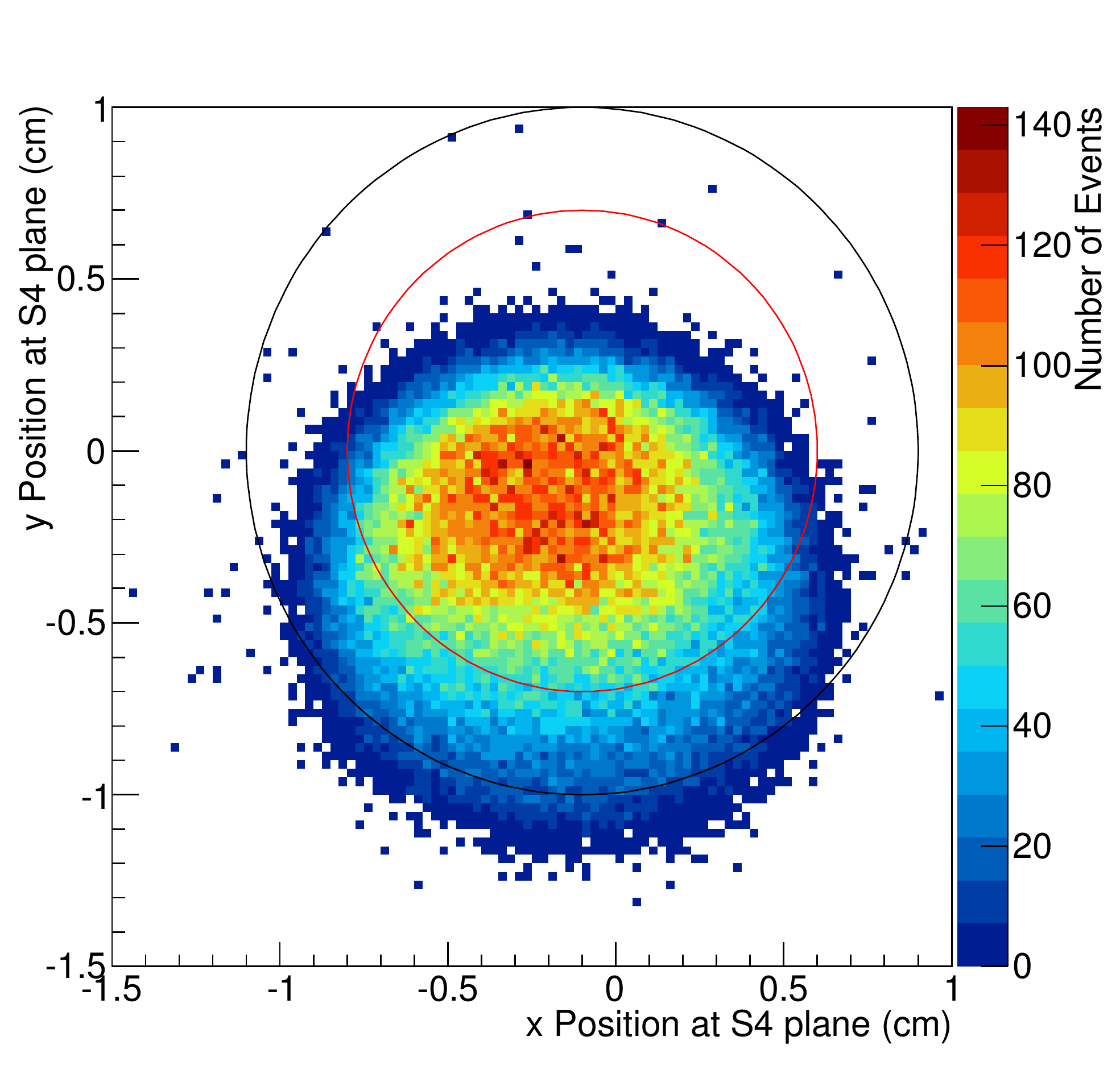}
\includegraphics[width=0.45\textwidth]{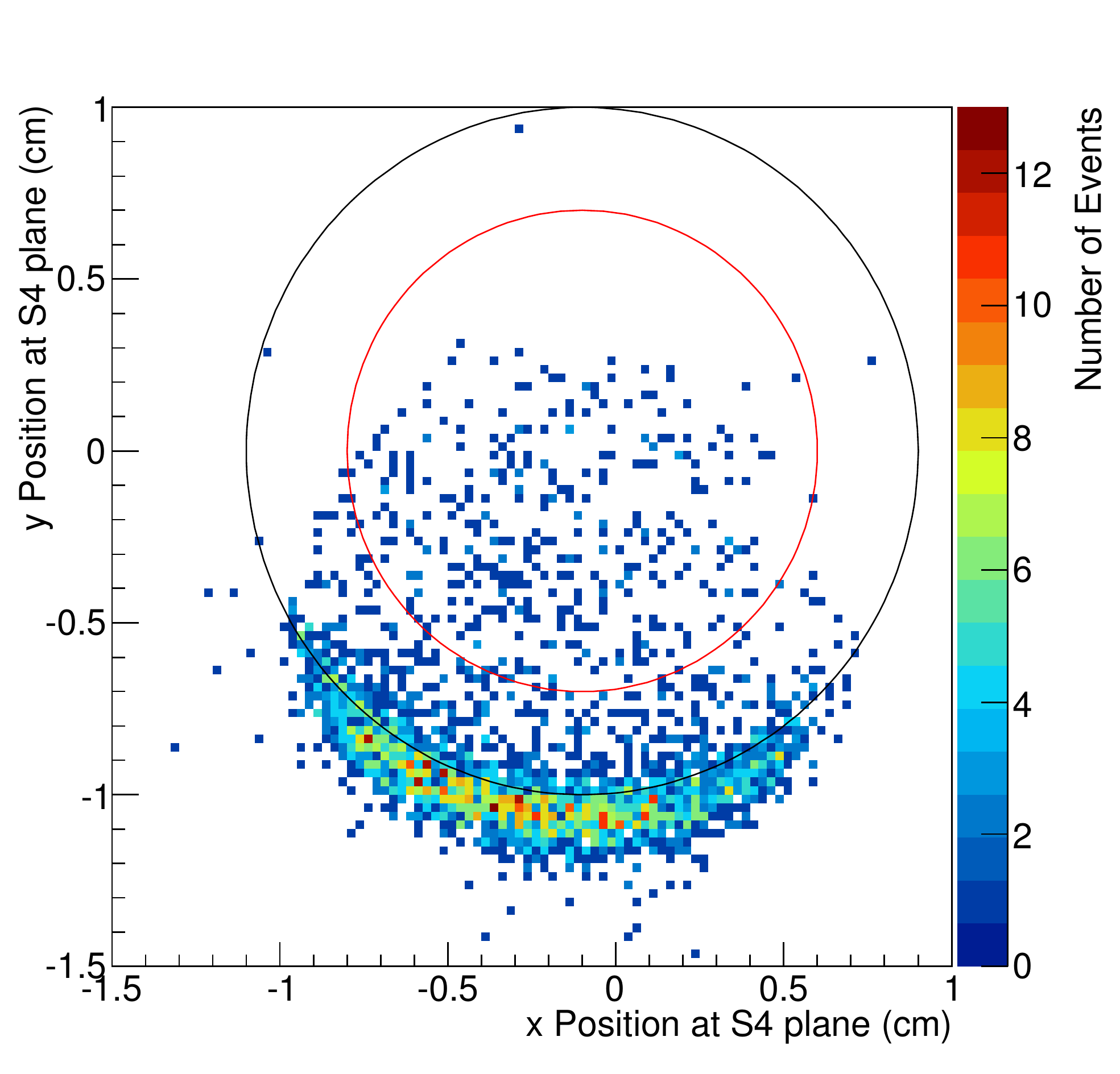}
\caption{Positions of BPD tracks extrapolated to the S4 plane in Target Removed data runs from the $\pi^{+} + \mbox{C}$ at 31\,\gevc dataset. 
The measured S4 position is shown as a black circle and the BPD radius cut is shown as a red circle in both figures.
($Left$) Events taken by the beam trigger.  
($Right$) Events taken by the interaction trigger.}
\label{fig:bpdExtrap31}
\end{center}
\end{figure*}

\begin{figure*}[tb]
\begin{center}
\includegraphics[width=0.45\textwidth]{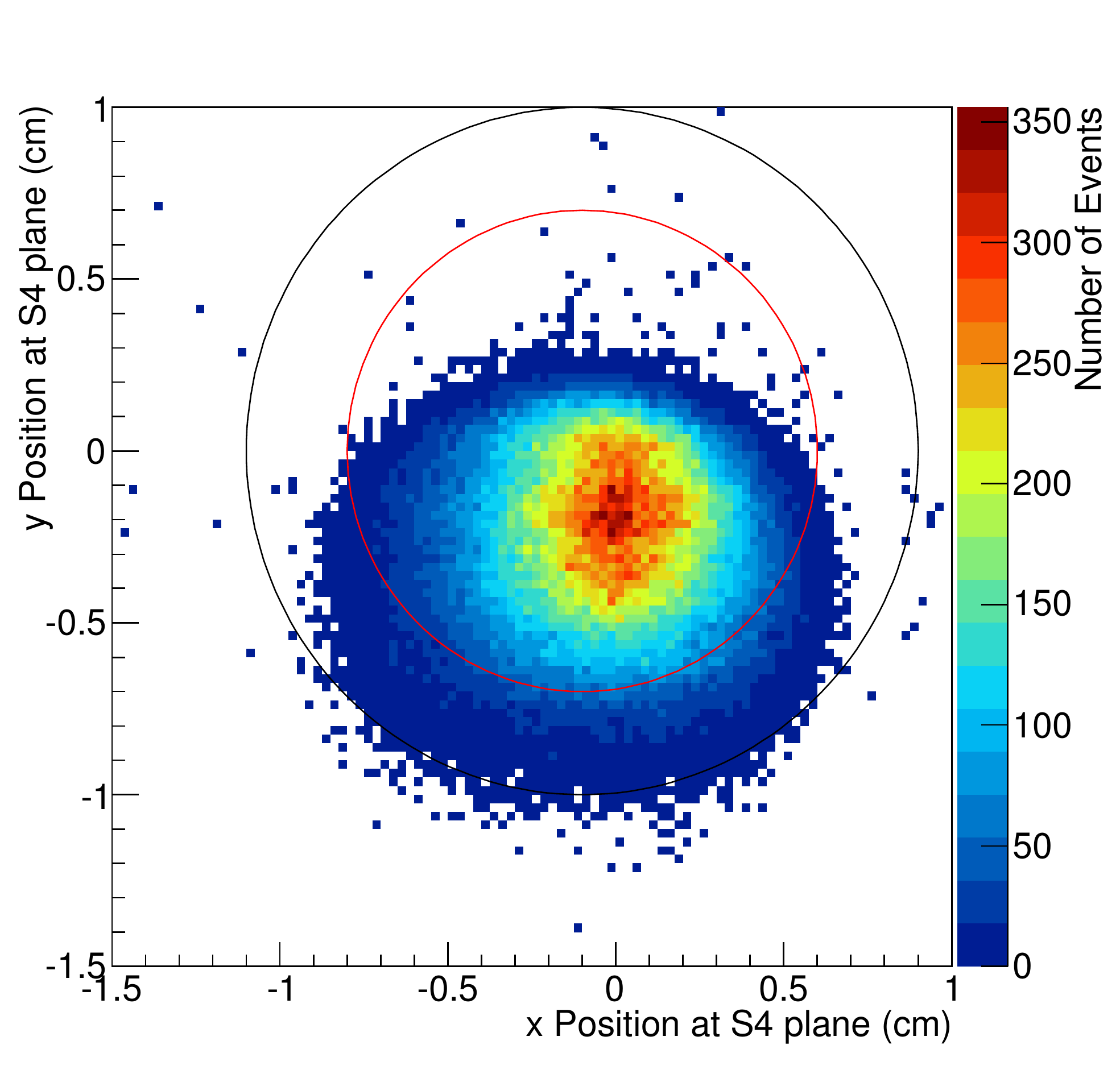}
\includegraphics[width=0.45\textwidth]{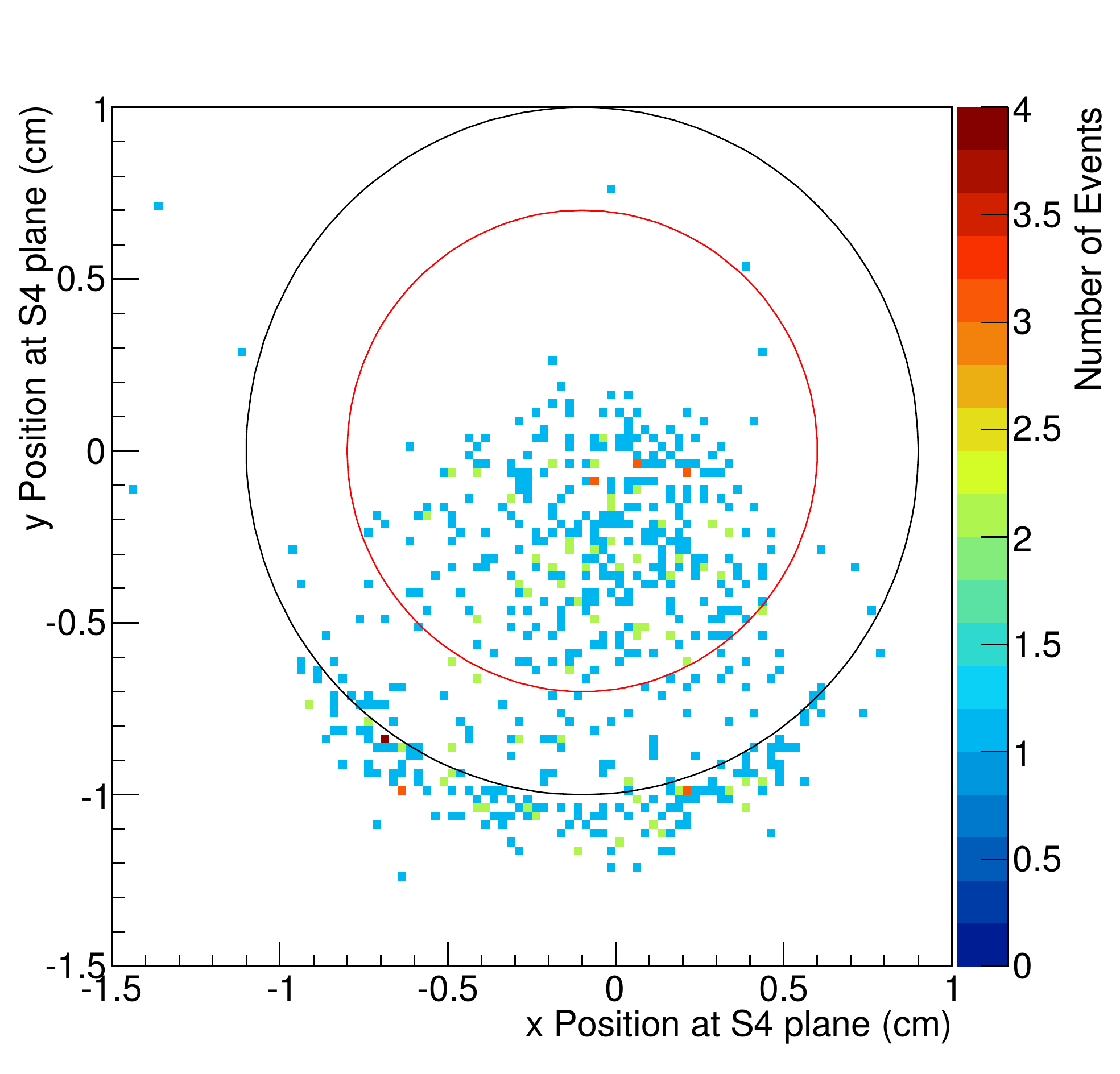}
\caption{Positions of BPD tracks extrapolated to the S4 plane in Target Removed data runs from the $\pi^{+} + \mbox{C}$ at 60\,\gevc dataset. 
The measured S4 position is shown as a black circle and the BPD radius cut is shown as a red circle in both figures.
($Left$) Events taken by the beam trigger.
($Right$) Events taken by the interaction trigger.}
\label{fig:bpdExtrap60}
\end{center}
\end{figure*}

The trajectories of the incoming beam particles are measured by three BPDs, located along the beamline upstream of the target as shown in Figure~\ref{fig:exp_setup}.
The measurements from the BPDs are especially important for estimating the effects of beam divergence on the cross section measurements. To understand these effects, tracks are fitted to the reconstructed BPD clusters, and the tracks are extrapolated to the S4 plane. The so-called ``Good BPD" cut requires that the event includes a cluster in the most-downstream BPD and that a track was successfully fit to the BPDs. Figures~\ref{fig:bpdExtrap31} and ~\ref{fig:bpdExtrap60} show the resulting BPD extrapolation to the S4 plane for the interactions studied.  The left plots show the extrapolated positions for all beam particles that pass the beam trigger, and the right plots the extrapolated positions for beam particles that pass the interaction trigger, which requires an anti-coincidence with the S4 scintillator. It can be seen from these figures that the 31\,\gevc beams were much wider than the 60\,\gevc beams. From these figures, it is also evident that the V1 veto counter (which is close to the most downstream BPD) and S4 were not well-aligned, especially for the 31\,\gevc beam. The beam was wide enough that a significant fraction of the beam particles have trajectories missing the S4, as can be see in the right plot of Figure~\ref{fig:bpdExtrap31} where a halo of beam particles miss the edge of the S4, mimicking the interaction trigger. This leads to an apparent interaction rate higher than the actual interaction rate. To reduce this effect, a radial cut was applied to the BPD tracks extrapolated to the S4, and this is indicated by the red circles on Figures~\ref{fig:bpdExtrap31} and ~\ref{fig:bpdExtrap60}.

The number of events after the described selection cuts for the interactions: 60\,\gevc $K^{+}$ and $\pi^{+}$ and 31\,\gevc $\pi^{+}$ with C and Al targets (Target Inserted) and with the targets removed (Target Removed) are shown in Tables 
[\ref{tab:event_pi31} - \ref{tab:event_k60}].

\begin{table}[tbh]
\centering
\begin{tabular}{ccccc}
Interaction & \multicolumn{2}{c}{$ \pi^{+} + \mbox{C}$} & \multicolumn{2}{c}{$ \pi^{+} + \mbox{Al}$} \\
\hline
Target & Inserted & Removed & Inserted & Removed \\
\hline
Total  & 593k & 195k & 535k & 234k \\[0.2ex]
WFA & 591k & 195k & 532k & 233k \\[0.2ex]
Good BPD  & 547k & 180k & 491k & 215k \\[0.2ex]
Radial cut  & 437k & 143k & 367k & 159k \\
\end{tabular}
\caption{ Event selection table for $ \pi^{+} + \mbox{C}$ and $ \pi^{+} + \mbox{Al}$ at 31\,\gevc.}
\label{tab:event_pi31}
\end{table}

\begin{table}[tbh]
\centering
\begin{tabular}{ccccc}
Interaction & \multicolumn{2}{c}{$ \pi^{+} + \mbox{C}$} & \multicolumn{2}{c}{$ \pi^{+} + \mbox{Al}$} \\
\hline
Target & Inserted & Removed & Inserted & Removed \\
\hline
Total            & 528k & 247k & 459k & 286k \\[0.2ex]
WFA             & 513k & 240k & 448k & 279k \\[0.2ex]
Good BPD    & 479k & 225k & 417k & 260k \\[0.2ex]
Radial cut    & 463k & 217k & 405k & 252k \\
\end{tabular}
\caption{ Event selection table for $ \pi^{+} + \mbox{C}$ and $ \pi^{+} + \mbox{Al}$ at 60\,\gevc.}
\label{tab:event_pi60}
\end{table}

\begin{table}[tbh]
\centering
\begin{tabular}{ccccc}
Interaction & \multicolumn{2}{c}{$ K^{+} + \mbox{C}$} & \multicolumn{2}{c}{$ K^{+} + \mbox{Al}$} \\
\hline
Target & Inserted & Removed & Inserted & Removed \\
\hline
Total          & 505k  & 239k & 339k & 156k \\[0.2ex]
WFA           & 503k  & 238k & 337k & 155k \\[0.2ex]
Good BPD  & 466k & 221k & 312k & 144k \\[0.2ex]
Radial cut  & 463k  & 219k & 310k  & 143k \\
\end{tabular}
\caption{Event selection table for $K^{+} + \mbox{C}$ and $ K^{+} + \mbox{Al}$ at 60\,\gevc.}
\label{tab:event_k60}
\end{table}

\section{Interaction trigger cross sections}\label{sec:trigger} \label{sec:ine_xsec}
In general, the probability of a beam particle interaction inside of a thin target is proportional to the thickness $L$ of the target and the number density of the target nuclei $n$. Thus, the interaction probability $P$ can be defined by taking into account the thin target approximation and by defining the interaction cross section $\sigma$ as:
\begin{eqnarray}\label{eq:p}
P = \frac{\text{ Number of events}}{\text{Number of beam particles}} = n\cdot L\cdot \sigma.
\end{eqnarray}
The density of nuclei $n$ can be calculated in terms of $N_{A}$, $\rho$, and $A$, which are Avogadro's number, the material density, and the atomic number, respectively.

\begin{figure*}[tb]
\begin{center}

\includegraphics[width=0.45\textwidth]{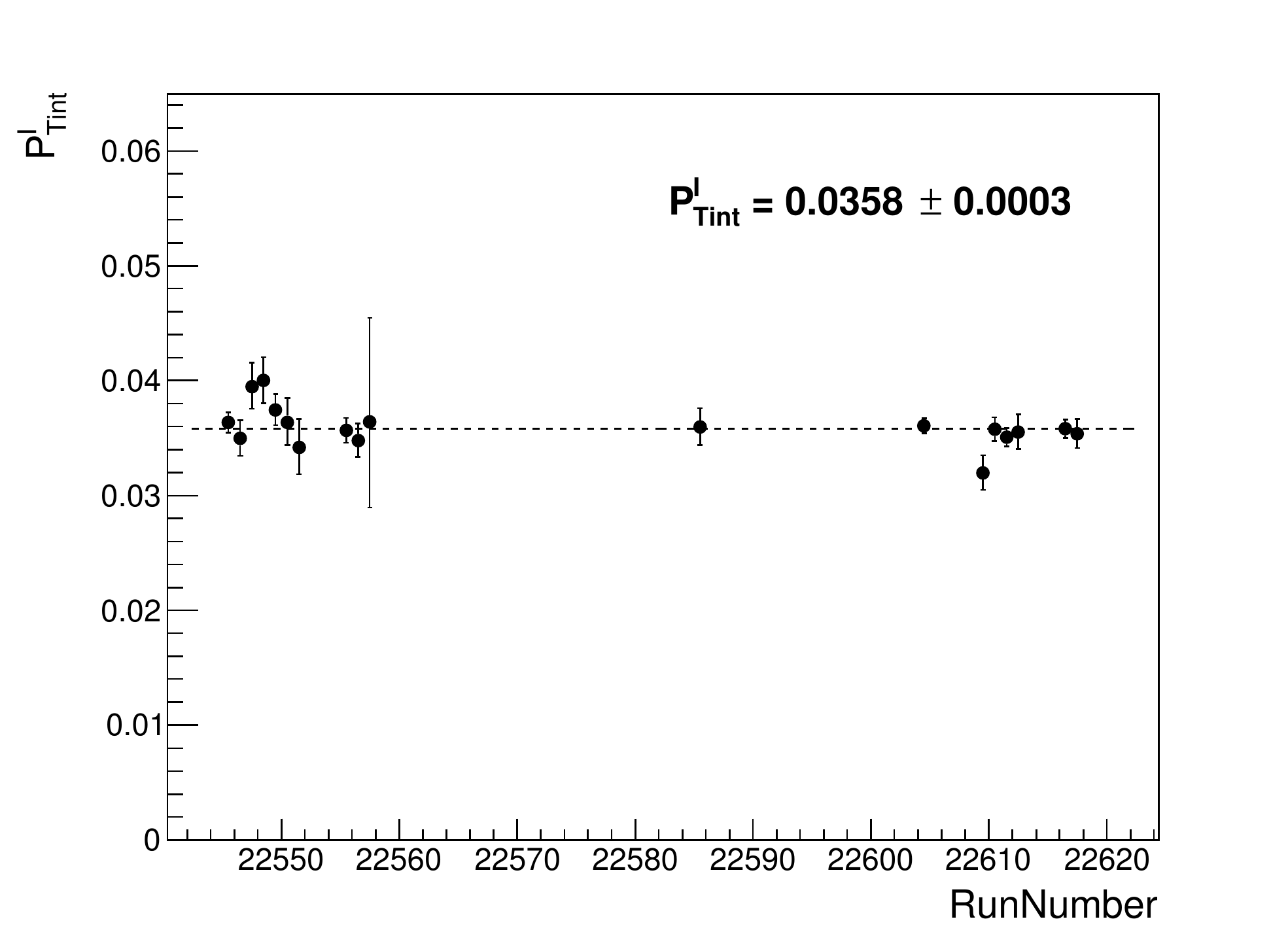}
\includegraphics[width=0.45\textwidth]{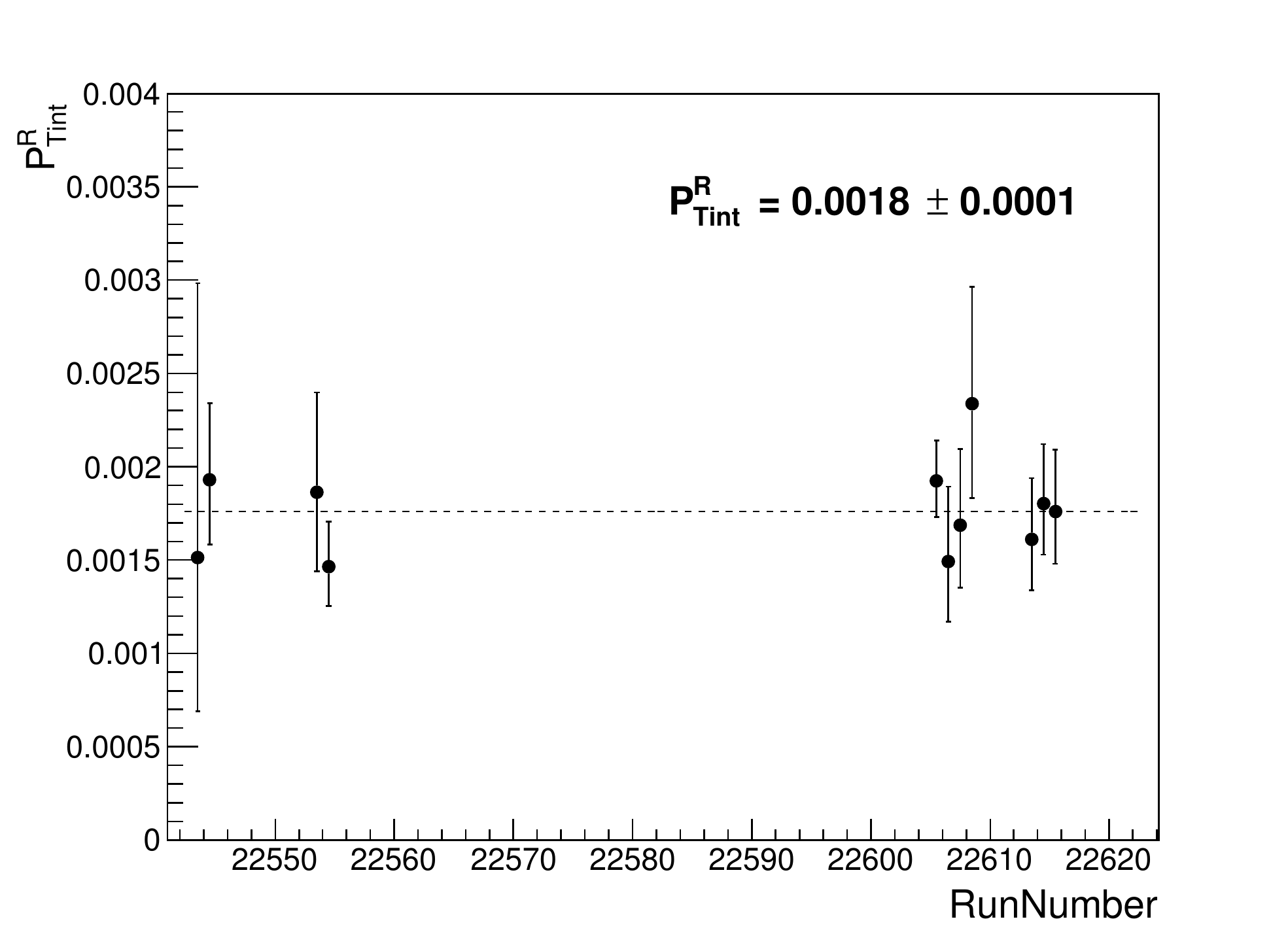}
\caption{Trigger interaction probabilities for the $\pi^{+} + \mbox{C}$ at 60 \gevc dataset.
($Left$) Target Inserted dataset. ($Right$) Target Removed dataset.}

\label{fig:ptint}
\end{center}
\end{figure*}

The counts of beam and interaction triggers as described in Sec.~\ref{sec:Setup} can be used to estimate the trigger probability as follows:
\begin{eqnarray}
 P_\mathrm{Tint} = \frac{N(T_\mathrm{beam}\wedge T_\mathrm{int})}{N(T_\mathrm{beam})}, \label{eq:P_Tint}
\end{eqnarray}
where $N(T_{beam})$ is the number of beam events passing the event selection cuts and $N(T_{beam}\wedge T_{int})$ is the number of selected beam events which also have an interaction trigger. 
In order to correct for events in which the beam particle interacts outside of the target, data were also recorded with the target removed from the beam (Target Removed) by rotating the target holder out of the path of the beam.  Figure~\ref{fig:ptint} shows an example of the trigger interaction probabilities  for each run for the $\pi^{+} + \mbox{C}$ at 60 \gevc dataset. Table~\ref{tab:ptint} gives the total trigger interaction probabilities for the data sets used in this paper for both the Target Inserted and Target Removed data.  The kaon target removed interaction probabilities are larger than those for pions due to the fact that $\sim$1\% of the beam kaons will decay between BPD\,3 and S4. 

Taking into account the trigger probabilities with the target inserted (\emph{I}) and the target removed (\emph{R}), $P_\mathrm{Tint}^\mathrm{I}$ and $P_\mathrm{Tint}^\mathrm{R}$, the interaction probability $P_\mathrm{int}$ can be obtained:
\begin{eqnarray}
 P_\mathrm{int} = \frac{P_\mathrm{Tint}^\mathrm{I} - P_\mathrm{Tint}^\mathrm{R}}{1 - P_\mathrm{Tint}^\mathrm{R}}. \label{eq:Pint}
\end{eqnarray}
Equation~\ref{eq:p} leads to the definition of the trigger cross section $\sigma_\mathrm{trig}$, calculated with $ P_\mathrm{int} $ and the effective target length $L_\mathrm{eff}$, which accounts for the exponential beam attenuation:
\begin{eqnarray}
 \sigma_\mathrm{trig} = \frac{A}{\rho L_\mathrm{eff} N_\mathrm{A}}\cdot P_\mathrm{int}.\label{eq:sigma}
\end{eqnarray}
The effective target length can be calculated with the absorption length,
\begin{eqnarray}
L_\mathrm{eff} = \lambda_\mathrm{abs}(1 - e^{-L/\lambda_\mathrm{abs}}),\label{eq:Leff}
\end{eqnarray}
where
\begin{eqnarray}
\lambda_\mathrm{abs} = A/(\rho N_\mathrm{A} \sigma_\mathrm{trig}).\label{eq:lam}
\end{eqnarray}
By combining Equations~\ref{eq:sigma}, ~\ref{eq:Leff}, and ~\ref{eq:lam}, one can simplify the equation for $\sigma_\mathrm{trig}$ as
\begin{eqnarray}
 \sigma_\mathrm{trig} = -\frac{A}{\rho L N_\mathrm{A}} \text{ln}(1-P_\mathrm{int}).\label{eq:sigma1}
\end{eqnarray}

\begin{table*}[tbh]
\centering
\begin{tabular}{cccccccc}
Interaction  & $p\,(\gevc) $ & $P_\mathrm{Tint}^\mathrm{I}$ & $P_\mathrm{Tint}^\mathrm{R}$ \\
\hline
$ \pi^{+} + \mbox{C}$&  31 & 0.0407 $\pm$ 0.0003 & 0.0025 $\pm$ 0.0001 \\
$ \pi^{+} + \mbox{Al}$& 31 & 0.0391 $\pm$ 0.0003 & 0.0029 $\pm$ 0.0001 \\
$ \pi^{+} + \mbox{C}$ & 60 & 0.0358 $\pm$ 0.0003 & 0.0018 $\pm$ 0.0001 \\
$ \pi^{+} + \mbox{Al}$& 60 & 0.0320 $\pm$ 0.0003 & 0.0018 $\pm$ 0.0001 \\
$ K^{+} + \mbox{C}$& 60 & 0.0394 $\pm$ 0.0003 & 0.0103 $\pm$ 0.0002\\
$ K^{+} + \mbox{Al}$& 60 & 0.0373 $\pm$ 0.0004 & 0.0103 $\pm$ 0.0003\\
\end{tabular}
\caption{Trigger Interaction probabilities in data. For each configuration, the observed probabilities for Target Inserted and Target Removed data are given.} 
\label{tab:ptint}
\end{table*}

\section{Correction factors}
\label{sec:corr_factors}
\subsection{S4 trigger correction factors}\label{sec:s4_xsec}

The trigger cross section contains the interactions where the resulting particles miss the S4 scintillator counter that is downstream of the target.  But even when there has been an interaction in the target, there is a possibility that a forward-going particle will strike the S4 counter. Moreover, not all elastically scattered beam particles strike the S4. Corrections must be applied to account for these effects.  Combining Equations~\ref{eq:tot_xsec} and~\ref{eq:prod_xsec}, the trigger cross section can be related to the production cross section through Monte Carlo (MC) correction factors as follows:
\begin{equation}
\sigma_\mathrm{trig}=\sigma_\mathrm{prod}\cdot f_\mathrm{prod}+\sigma_\mathrm{qe}\cdot f_\mathrm{qe}+\sigma_\mathrm{el}\cdot f_\mathrm{el}\ ,\label{eq:s4_fact}
\end{equation}
where $f_\mathrm{prod}$, $f_\mathrm{qe}$, and $f_\mathrm{el}$ are the fractions of production, quasi-elastic, and elastic events that miss the S4 counter. $\sigma_\mathrm{qe}$ and $\sigma_\mathrm{el}$ are also estimated from Monte Carlo.  
Equation~\ref{eq:s4_fact} can be rewritten to obtain $\sigma_\mathrm{prod}$ and $\sigma_\mathrm{inel}$ as:
\begin{equation}
\sigma_\mathrm{prod}= \frac{1}{f_\mathrm{prod}}( \sigma_\mathrm{trig}  - \sigma_\mathrm{qe}\cdot f_\mathrm{qe} - \sigma_\mathrm{el}\cdot f_\mathrm{el})\label{eq:sig_prod}
\end{equation}
and
\begin{equation}
\sigma_\mathrm{inel}= \frac{1}{f_\mathrm{inel}}( \sigma_\mathrm{trig}  -  \sigma_\mathrm{el}\cdot f_\mathrm{el}).\label{eq:sig_inel}
\end{equation}


A GEANT4  detector simulation~\cite{Agostinelli:2002hh, Allison:2006ve, Allison:2016lfl} was used to estimate the MC correction factors discussed above.
The FTFP\_BERT physics list with GEANT4 version of 10.2.p03 was used to estimate correction factors as presented in Table~\ref{tab:MC_factors}.

\begin{table*}[tbh]
\centering
\begin{tabular}{cccccccc}
Interaction  & $p$ & \multicolumn{6}{c}{Monte Carlo Correction Factors}\\
\cline{3-8}
                    & (\gevc) & $\sigma_\mathrm{el}$ (mb)& $f_\mathrm{el}$ & $\sigma_\mathrm{qe}$ (mb) & $f_\mathrm{qe}$ & $f_\mathrm{prod}$ & $f_\mathrm{inel}$\\ 

\hline
$ \pi^{+} + \mbox{C}$&  31 & 55.5 & 0.734 & 18.8 & 0.946 & 0.989 & 0.985\\
$ \pi^{+} + \mbox{Al}$& 31 & 114.5& 0.745 & 29.7 & 0.949 & 0.990 & 0.987\\
$ \pi^{+} + \mbox{C}$ &  60 & 54.0 & 0.289 & 16.4& 0.811 & 0.967& 0.952\\
$ \pi^{+} + \mbox{Al}$& 60 & 110.0 & 0.232 & 25.7& 0.814 & 0.969 & 0.956\\
$ K^{+} + \mbox{C}$&    60 & 18.1 & 0.323 & 14.5 & 0.821 & 0.990 & 0.975\\
$ K^{+} + \mbox{Al}$&   60 & 44.6 & 0.183 & 23.5 & 0.821 & 0.990 & 0.997\\
\end{tabular}
\caption{Monte Carlo correction factors. 
} 
\label{tab:MC_factors}
\end{table*}

\subsection{Beam composition correction factors}\label{sec:beamcomp_corr}

In the case of \pip beams, a correction must also be applied to account for contamination from \mup and \ePl .  The CEDAR and threshold Cherenkov detectors do not have the power to completely discriminate positrons and muons from pions at 31\,\gevc and 60\,\gevc as shown in ~\cite{cedar, cedar82}. Fortunately, it was possible to estimate the amount of positron contamination with the TPC system and the PSD. During the neutrino data-taking in 2016, a special maximum field data run was taken during which the 60\,\gevc \hp beam was bent into the MTPC-L.  

\begin{figure*}[tb]
\begin{center}
\includegraphics[width=0.55\textwidth]{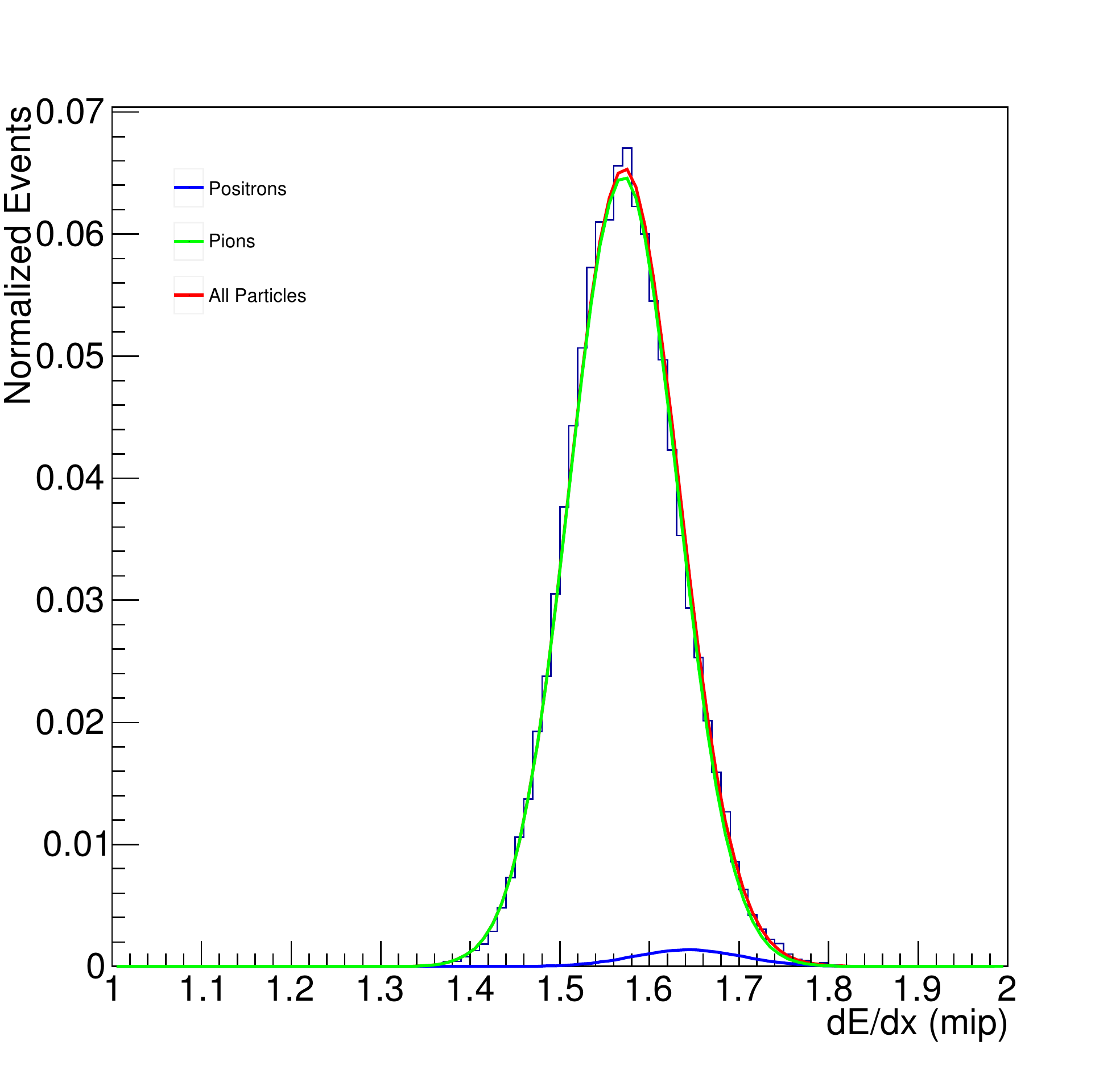}
\caption{The binned data shows the $dE/dx$ distribution of the maximum field dataset for the 60\,\gevc $\pi^+$ beam. Overlaid is the sum of gaussians fit to the histogram as well as the individual $\pi^{+}$ and $e^{+}$ components. From this fit, the positron contribution was estimated to be 2\%.}
\label{fig:maxFieldFit}
\end{center}
\end{figure*}

Positrons deposit most of their energy in the first 2 out of 10 longitudinal sections of the PSD, while pions penetrate deeper. A pure pion sample is obtained by selecting beam particles that deposit less than 20\% of their total energy in the first section of the PSD.  To determine the \ePl and \pip compositions of the beam, a sum of two gaussians is fit to the \dedx distribution.
From the fit, the positron contamination was determined to be 2\%$\pm$2\% for the 60\,\gevc beam. Figure~\ref{fig:maxFieldFit} shows the resulting fit to the maximum field data.  The GEANT4 MC simulation is used to determine the effect of the positrons on the trigger cross section.  
The resulting corrections applied to $\sigma_\mathrm{prod}$ ($\sigma_\mathrm{inel}$) are +2.2\% (+2.1\%)
for $\pi^{+} + \mbox{C}$ at 60\,\gevc and +1.8\% (1.7\%) for $\pi^{+} + \mbox{Al}$ at 60\,\gevc .

In the case of 31\,\gevc,  the potential for positron contamination was reduced by requiring that the CEDAR had a more stringent 7-fold coincidence signal. No special data run was undertaken with the 31 \,\gevc beam to measure the positron contamination, so no correction is applied. But this contamination will be taken into account later as an asymmetric systematic uncertainty. 

For the pion beams at both 31\,\gevc and 60\,\gevc, a small number of muons are also present in the beam due to the decays of pions upstream of the target.  Many of these muons diverge from the beam and will strike the veto counters, but beam simulations at both momenta show the muon fraction that will pass the veto counters and trigger our beam counters is about 1.5$\pm$0.5\% of the pion beam.  A correction for the muon component of the beam is applied to the 31\,\gevc and 60\,\gevc pion beam interactions.  

For the kaon beam, any kaons that decay upstream of the CEDAR will not satisfy the beam selection and will not be selected as good beam particles.  Only kaon decays downstream of the CEDAR where the decay products head toward the S4 will pass the beam selection and ``Good BPD" cut.  It was estimated that only  0.1\% of the CEDAR-tagged kaons will decay with decay products that pass these cuts. Therefore, no correction is applied for kaon decays in the beamline.

\section{Systematic uncertainties}\label{sec:systematics}

\subsection{Target density uncertainty}

The uncertainty on the target density affects the calculation of the trigger cross section as shown in Equation~\ref{eq:sigma1}. The density uncertainty for each target is estimated by calculating the standard deviation of the target densities determined from measurements of the mass and dimensions of the machined target samples. A 0.65\% uncertainty on the density of carbon and a 0.29\% uncertainty on the density of aluminum were used. 

\subsection{Out-of-target interactions}

As shown in Equation~\ref{eq:Pint}, the measured interaction rates are corrected for interactions occurring outside of the target by measuring the trigger rates with the target both inserted and removed. To look for possible additional systematic effects, two special runs were undertaken with the target holder in the ``I'' position and with the target holder in the ``R'' position, but with no target attached. The data were taken with 31\,\gevc and 60\,\gevc $\pi^{+}$. 

In the case of the 31\,\gevc target holder data, there was no significant difference between the trigger probability of the empty target holder data and the target removed data.
However, in the case of the 60\,\gevc data, a high trigger probability in the target holder ``I'' run was observed. 
These additional out-of-target interactions may be related to the beam conditions during those runs. An asymmetric uncertainty was assigned for the 60\,\gevc interactions.  

\subsection{S4 size uncertainty and efficiency}

Another systematic uncertainty comes from the uncertainty in the size of the S4 scintillator. The diameter of the S4 has previously been found to have an uncertainty of  $\pm 0.40$ mm.  In order to propagate this uncertainty to $\sigma_\mathrm{inel}$ and $\sigma_\mathrm{prod}$,  two additional MC simulation samples with the S4 diameter modified 
were generated. 


Previous NA61/SHINE analyses have found that S4 inefficiency is negligibly small~\cite{Abgrall:2011ae} and this analysis also found no S4 inefficiency by looking at GTPC tracks in Target Removed data. The S4 inefficiency is concluded to be less than 0.1\% and neither an uncertainty nor a correction relating to the S4 scintillator efficiency is applied to the results.

\subsection{Beam composition uncertainty}

As was mentioned in Section~\ref{sec:beamcomp_corr}, for interactions with the 60\,\gevc $\pi^{+}$ beam, a correction was applied to reflect the small amount of positrons in the beam.  To be conservative, 100\% of this correction is assumed as a systematic uncertainty.   For $\pi^{+} $ at 31 \gevc, no correction is applied, but an uncertainty is reported accounting for a 1\% positron contamination.

As was also mentioned in Section~\ref{sec:beamcomp_corr}, the muon fraction in the pion beam is estimated to be 1.5\% for both the 31\,\gevc and 60\,\gevc $\pi^{+}$ beams and a correction was applied. An uncertainty of 0.5\% is applied to this correction.

The CEDAR counter has a high purity of identifying kaons using a 6-fold coincidence for 60\,\gevc beams. The lower limit on the purity of the kaon beam was calculated to be 99.4\% according to the CEDAR gas pressure scan data. The estimated systematic error from this source is applied to the total systematic uncertainty. 


\subsection{Model uncertainties}

The S4 correction factors $f_\mathrm{prod}$, $f_\mathrm{inel}$, $f_\mathrm{el}$ and $f_\mathrm{qe}$ as well as the cross sections $\sigma_\mathrm{qe}$ and $\sigma_\mathrm{el}$ were estimated with GEANT4 MC simulations using the FTFP\_BERT physics list. In order to estimate the model uncertainties associated with these correction factors, the correction factors were recalculated with three additional physics lists: QBBC, QGSP\_BERT and FTF\_BIC. 
Using these additional physics lists, the model dependency on the total cross section measurements was studied. 


\begin{table*}[tbh]
\centering
\begin{tabular}{ccccccccc}
  &  & \multicolumn{5}{c}{Systematic uncertainties for $\sigma_\mathrm{prod}$ (mb)}& & \\
\cline{3-7}
                    &       $p$      &            & Out-of- & S4  & Beam  & MC  & Total Syst. & Model\\ 
    Interaction  & (\gevc) & Density & target & Size  &Purity & Stat. & Uncer. & Uncer.\\ 

\hline
$ \pi^{+} + \mbox{C}$&  31 & $\pm 1.4$ &--& $\pm ^{0.9} _{0.7}$ & $\pm ^{2.3} _{1.1}$   & $\pm 0.3$ & $\pm ^{2.8} _{2.0}$ & $ \pm ^{1.1} _{0.4}$ \\[0.2ex]
$ \pi^{+} + \mbox{Al}$& 31 & $\pm 1.2$ &-- & $ \pm ^{1.8} _{1.8}$ &  $\pm ^{3.5} _{2.2}$  & $\pm 0.6$ & $\pm ^{4.2} _{3.1}$ & $ \pm ^{3.9} _{0.6}$ \\[0.2ex]
$ \pi^{+} + \mbox{C}$&  60 & $\pm 1.3$ & $\pm ^{0.0} _{1.2} $& $ \pm ^{1.4} _{1.3}$ & $\pm ^{4.0} _{3.8}$ & $\pm 0.3$ & $\pm ^{4.4}_{4.4}$ & $\pm ^{0.4} _{1.4}$\\[0.2ex]
$ \pi^{+} + \mbox{Al}$& 60 & $\pm 1.1$ & $\pm^{0.0}_{4.3}$ & $\pm ^{2.4}_{2.8}$ & $\pm^{6.4}_{6.1}$  & $\pm 0.6$ & $\pm  ^{6.9} _{8.1}$ & $\pm ^{0.8}_{0.7}$\\[0.2ex]
$ K^{+} + \mbox{C}$&    60 & $\pm 0.8$  & $\pm 0.6$ & $\pm ^{0.3}_{0.3}$ & $\pm^{0.3}_{0.3}$  & $\pm 0.1$ &  $\pm ^{1.1}_{1.1}$ & $\pm ^{0.2}_{2.9}$\\[0.2ex]
$ K^{+} + \mbox{Al}$&   60 & $\pm 1.1$  & $\pm 1.2$ & $\pm ^{0.5} _{0.5}$ & $\pm^{0.5}_{0.5}$  & $\pm 0.1$ &  $\pm ^{1.8}_{1.8}$ & $\pm ^{0.1}_{4.1}$\\
\end{tabular}
\caption{Breakdown of systematic uncertainties for production cross section measurements with the NA61/SHINE data. 
} 
\label{tab:sys_prod}
\end{table*}

\begin{table*}[tbh]
\centering
\begin{tabular}{ccccccccc}
       &  & \multicolumn{5}{c}{Systematic uncertainties for $\sigma_\mathrm{inel}$ (mb)}& & \\
\cline{3-7}
                    &       $p$      &            & Out-of- & S4  & Beam  & MC  & Total Syst. & Model\\ 
    Interaction  & (\gevc) & Density & target & Size  &Purity & Stat. & Uncer. & Uncer.\\ 

\hline
$\pi^{+} + \mbox{C}$ & 31 & $\pm 1.4$  & --& $\pm^{0.9} _{0.7}$ & $\pm^{2.3} _{1.1}$& $\pm 0.3$  & $\pm^{2.8} _{2.0}$ & $\pm^{1.2} _{0.4}$ \\[0.2ex]
$ \pi^{+} + \mbox{Al}$& 31 & $\pm 1.2$ & --& $\pm^{1.8} _{1.8}$ & $\pm^{3.6} _{2.2}$& $\pm 0.6$   & $\pm^{4.2} _{3.2}$ & $\pm^{4.0} _{0.6}$ \\[0.2ex]
$ \pi^{+} + \mbox{C}$ & 60 & $\pm 1.3$ & $\pm^{0.0} _{1.3}$& $\pm^{1.4} _{1.2}$ & $\pm^{4.1} _{4.0}$& $\pm 0.3$   & $\pm^{4.5} _{4.6}$ & $\pm^{0.3} _{3.9}$\\[0.2ex]
$ \pi^{+} + \mbox{Al}$& 60 & $\pm 1.1$ & $\pm^{0.0} _{4.3}$& $\pm^{2.5} _{2.8}$ & $\pm^{6.4} _{6.2}$& $\pm 0.6$   & $\pm^{7.0} _{8.1}$ & $\pm^{1.1} _{0.8}$\\[0.2ex]
$ K^{+} + \mbox{C}$  & 60 &  $\pm 0.8$  & $\pm 0.6$ & $\pm^{0.3}_{0.4}$ & $\pm^{0.3}_{0.3}$ & $\pm 0.1$ & $\pm^{1.1} _{1.1}$ & $\pm^{0.1} _{2.3}$\\[0.2ex]
$ K^{+} + \mbox{Al}$ & 60 &  $\pm 1.1$  & $\pm 1.2$ & $\pm^{0.6}_{0.5}$ & $\pm^{0.5}_{0.5}$ & $\pm 0.1$ & $\pm^{1.8} _{1.8}$ & $\pm^{0.1} _{3.1}$\\
\end{tabular}
\caption{Breakdown of systematic uncertainties for inelastic cross section measurements with the NA61/SHINE data. 
} 
\label{tab:sys_inel}
\end{table*}

\section{Results}\label{sec:results}

\begin{figure*}[tb]
\begin{center}
\includegraphics[width=0.85\textwidth]{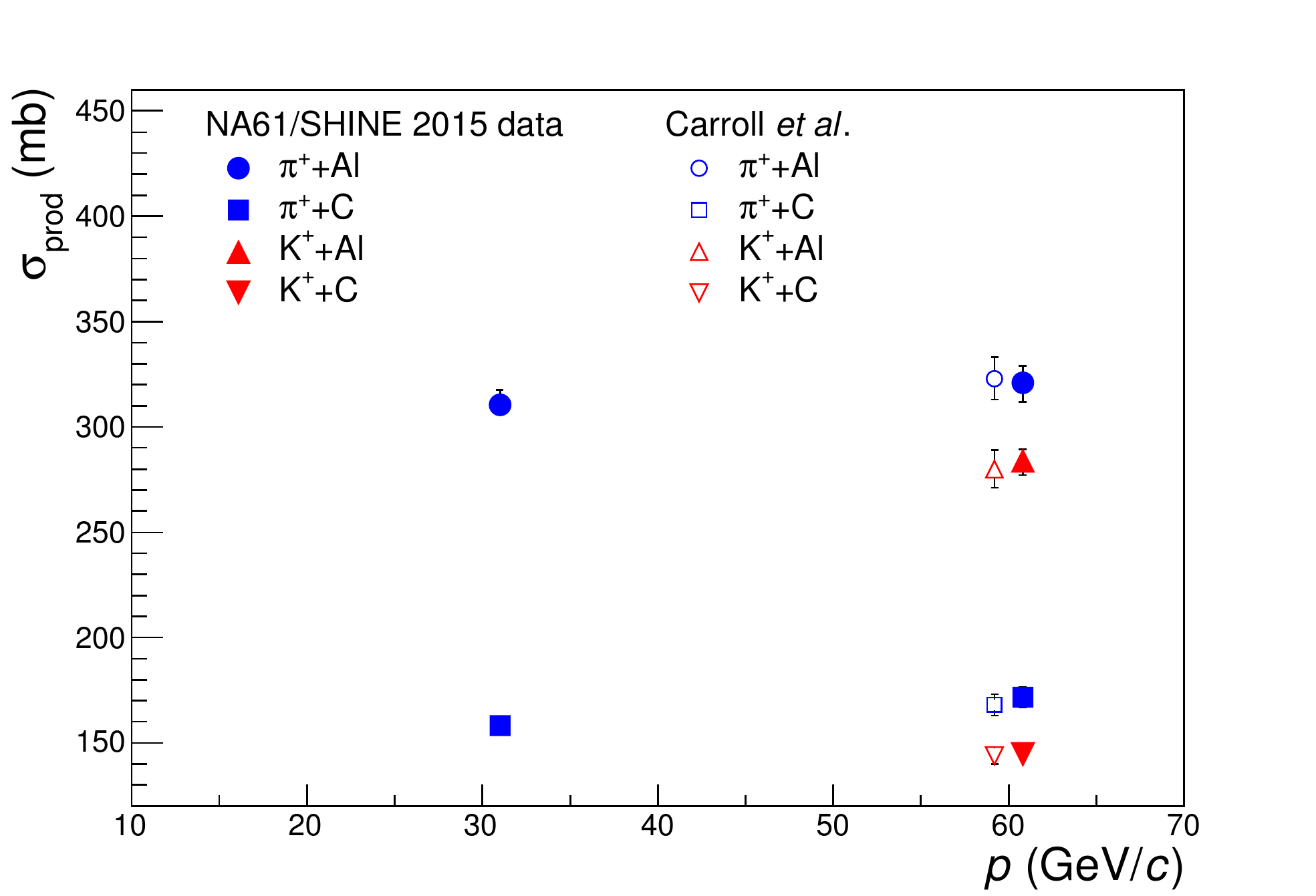}
\caption{Summary of production cross section measurements.
The results are compared to previous results obtained with a beam momentum of 60\,\gevc by Carrol \textit{et al}.~\cite{Carroll} . 
}
\label{fig:ProdXsec}
\end{center}
\end{figure*}

Several production cross sections have been measured in this analysis:
$\pi^++\mbox{C}$ ($\pi^++\mbox{Al}$) at 31\,\gevc is found to be 158.3\,mb (310.4\,mb),
$\pi^++\mbox{C}$ ($\pi^++\mbox{Al}$) at 60\,\gevc is found to be 171.6\,mb (321.0\,mb), and
$K^++\mbox{C}$ ($K^++\mbox{Al}$) at 60\,\gevc is found to be 144.5\,mb (284.0\,mb), respectively. 
Statistical, systematic, and physics model uncertainties are estimated separately and are summarized in Table~\ref{t:nuprodcross}.
$\pi^+$ and $K^+$ at 60\,\gevc measurements are compared with the results of Carrol \textit{et al}.~\cite{Carroll} 
as shown in Figure~\ref{fig:ProdXsec}.  These NA61 results are consistent within our errors with the previous measurements, and our error bands are smaller, especially for the kaons.

\begin{figure*}[tb]
\begin{center}
\includegraphics[width=0.85\textwidth]{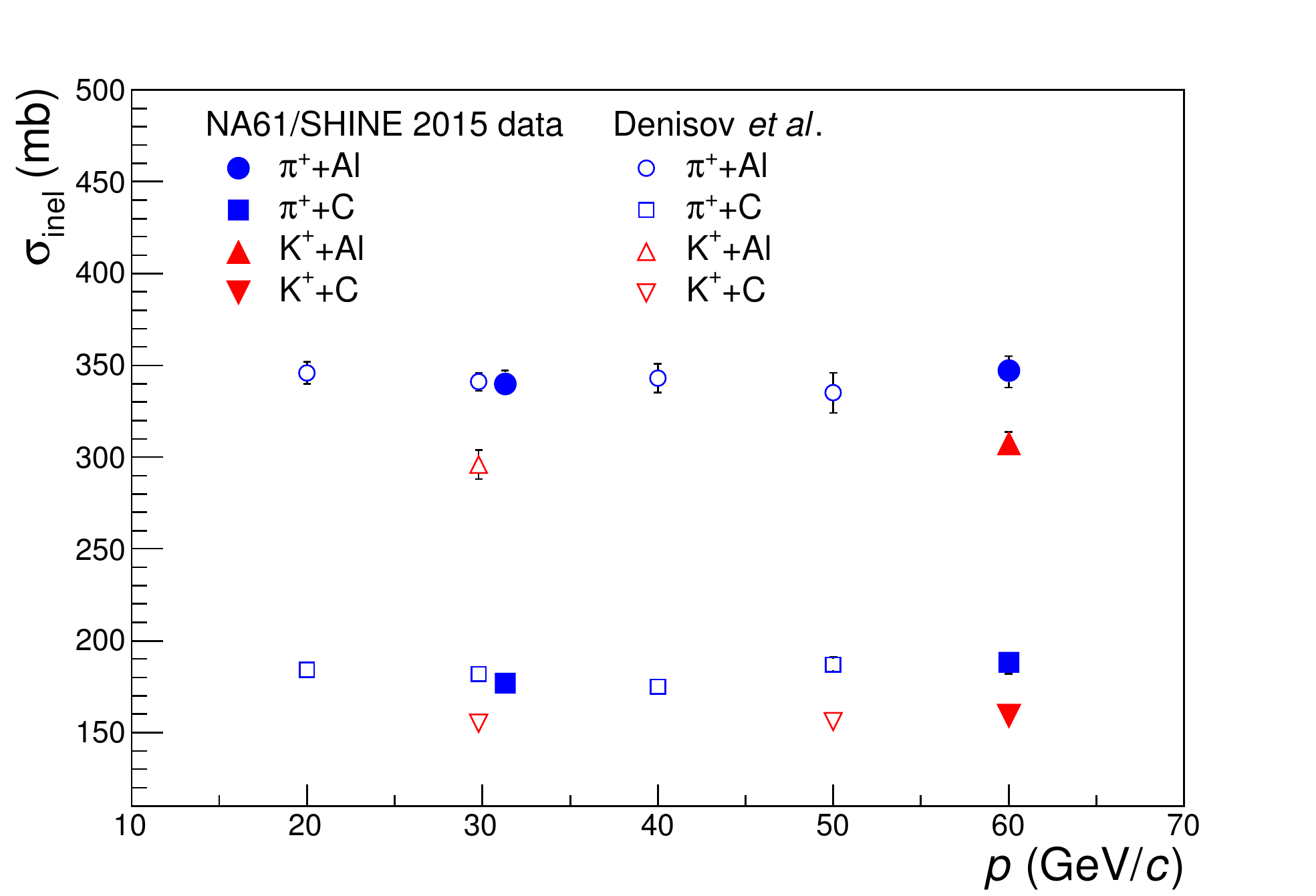}
\caption{Summary of inelastic cross section measurements.
The results are compared to previous results by Denisov \textit{et al}.~\cite{Denisov:1973zv} .
}
\label{fig:InelXsec}
\end{center}
\end{figure*}

Several inelastic cross sections have also been determined in this analysis: 
$\pi^++\mbox{C}$ ($\pi^++\mbox{Al}$) at 31\,\gevc is found to be 177.0\,mb (340.0\,mb),
$\pi^++\mbox{C}$ ($\pi^++\mbox{Al}$) at 60\,\gevc is found to be 188.2\,mb (347.0\,mb), and
$K^++\mbox{C}$ ($K^++\mbox{Al}$) at 60\,\gevc is found to be 159.0\,mb (307.5\,mb), respectively.
Statistical, systematic, and physics model uncertainties are estimated separately and are summarized in Table~\ref{t:nuinelcross}.
These measurements are compared with the results of Denisov \textit{et al}.~\cite{Denisov:1973zv}
as shown in Figure~\ref{fig:InelXsec}. These NA61 results are consistent within errors with the existing measurements at 30\,\gevc .

Additionally, a short data run of interactions of 31\,\gevc protons with carbon was analyzed as a cross-check with the previous  higher statistics NA61/SHINE total cross section results from the 2009 T2K data run~\cite{na61_t2k_thin}. The total production (total inelastic) cross section was found to be 229.8$\pm$ 4.4\,mb (259.9$\pm$4.5\,mb) (statistical uncertainty only). These are consistent with the 2009 result of 230.7\,mb (258.4\,mb).

\begin{table}[tbh]
\centering
\begin{tabular}{ccccccc}
Interaction  & $p$& \multicolumn{4}{c}{Production cross section (mb)} \\ 
                     & (\gevc)  &$\sigma_\mathrm{prod}$ & $\Delta_\mathrm{stat}$& $\Delta_\mathrm{syst}$ & $\Delta_\mathrm{model}$ & $\Delta_\mathrm{total}$ \\
\hline
$ \pi^{+} + \mbox{C}$ & 31 & 158.3 & $\pm 2.0$ & $\pm^{2.8} _{2.0}$ & $\pm^{1.1} _{0.4}$  & $\pm^{3.6} _{2.9}$ \\[0.2ex]
$ \pi^{+} + \mbox{Al}$& 31 & 310.4 & $\pm 4.3$ & $\pm^{4.2} _{3.1}$ & $\pm^{3.9} _{0.6}$ & $\pm^{7.2} _{5.3}$ \\[0.2ex]
$ \pi^{+} + \mbox{C}$&  60 & 171.6 & $\pm 1.7$ & $\pm^{4.4} _{4.4}$ & $\pm^{0.4} _{1.4}$ &$\pm^{4.7} _{4.9}$ \\[0.2ex]
$ \pi^{+} + \mbox{Al}$& 60 & 321.0 & $\pm 4.0$ & $\pm^{6.9} _{8.1}$ & $\pm^{0.8} _{0.7}$ & $\pm^{8.0} _{9.1}$\\[0.2ex]
$ K^{+} + \mbox{C}$ & 60  & 144.5 & $\pm 2.0$ & $\pm^{1.1} _{1.1}$ & $\pm^{0.2} _{2.9}$ & $\pm^{2.3} _{3.7}$ \\[0.2ex]
$ K^{+} + \mbox{Al}$& 60  & 284.0 & $\pm 5.1$ & $\pm^{1.8} _{1.8}$ & $\pm^{0.1} _{4.1}$ & $\pm^{5.4} _{6.8}$\\[0.2ex]
\end{tabular}
\caption{Production cross section measurements with the NA61/SHINE data. The central value as well as the statistical ($\Delta_\mathrm{stat}$), systematic ($\Delta_\textrm{syst}$), and model ($\Delta_\mathrm{model}$) uncertainties are shown.  The total uncertainty ($\Delta_\textrm{total}$) is the sum of the statistical, systematic, and model uncertainties in quadrature.}
\label{t:nuprodcross}
\end{table}

\begin{table}[tbh]
\centering
\begin{tabular}{ccccccc}
Interaction  & $p$ & \multicolumn{4}{c}{Inelastic cross section (mb)} \\ 
                     & (\gevc)  &$\sigma_\mathrm{inel}$ & $\Delta_\mathrm{stat}$& $\Delta_\mathrm{syst}$ & $\Delta_\mathrm{model}$ & $\Delta_\mathrm{total}$ \\
\hline
$ \pi^{+} + \mbox{C}$  & 31 &  177.0 & $\pm 2.0$ & $\pm^{2.8} _{2.0}$ & $\pm^{1.2} _{0.4}$ & $\pm^{3.6} _{2.9}$  \\[0.2ex]
$ \pi^{+} + \mbox{Al}$ & 31 & 340.0 & $\pm 4.4$ & $\pm^{4.2} _{3.2}$ &  $\pm^{4.0} _{0.6}$ & $\pm^{7.3} _{5.5}$ \\[0.2ex]
$ \pi^{+} + \mbox{C} $ & 60 &  188.2 & $\pm 1.8$ & $\pm^{4.5} _{4.6}$ & $\pm^{0.3} _{3.9}$ & $\pm^{4.9} _{6.3}$ \\[0.2ex]
$ \pi^{+} + \mbox{Al}$ & 60 &  347.0 & $\pm 4.1$ & $\pm^{7.0} _{8.1}$ & $\pm^{1.1} _{0.8}$ & $\pm^{8.2} _{9.1}$ \\[0.2ex]
$ K^{+} + \mbox{C} $   & 60  & 159.0  & $\pm 2.1$ & $\pm^{1.1} _{1.1}$ & $\pm^{0.1} _{2.3}$ & $\pm^{2.4} _{3.3}$ \\[0.2ex]
$ K^{+} + \mbox{Al} $  & 60 & 307.5  & $\pm 5.1 $ & $\pm^{1.8} _{1.8}$ & $\pm^{0.1} _{3.1}$ & $\pm^{5.4} _{6.2}$ \\[0.2ex]
\end{tabular}
\caption{Inelastic cross section measurements with the NA61/SHINE data. The central value as well as the statistical ($\Delta_\mathrm{stat}$), systematic ($\Delta_\textrm{syst}$), and model ($\Delta_\mathrm{model}$) uncertainties are shown. The total uncertainty ($\Delta_\textrm{total}$) is the sum of the statistical, systematic, and model uncertainties in quadrature.}
\label{t:nuinelcross}
\end{table}

\section{Summary}\label{sec:Discussion}
In summary, the production and inelastic cross sections of $\pi^+$ and $K^+$ on carbon and aluminum targets have been measured with the NA61/SHINE experiment.
The production cross section with $\pi^+$ beams at 31\,\gevc was measured for the first time with a precision of about 2\%. At
60\,\gevc the measured production cross sections are comparable to previous results for $\pi^+$ and $K^+$ and the precision was improved to about 3\% and 2\%, respectively.
Inelastic cross section measurements with $\pi^+$ and $K^+$ beams at 60\,\gevc were measured for first time with precisions of about 3\% and 2\%, respectively.
For the inelastic production cross section for $\pi^+$ at 31\,\gevc  reasonable agreement with a previous measurement was found.  Especially for $\pi^+$ beams, the measurements here are limited by positron contamination in the beam and steps will be taken  in future data-taking to better limit this uncertainty.

The current uncertainties on the neutrino fluxes in the NuMI neutrino beam at Fermilab from the MINER$\nu$A collaboration~\cite{numi_flux} rely on measurements of the inelastic cross section (which is termed the ``absorption" cross section in the  MINER$\nu$A paper).  For $\pi^{+}$+C and $\pi^{+}$+Al they assumed an uncertainty of 5\%, while for the $K^{+}$+C and $K^{+}$+Al cross sections they assumed a 10-30\% uncertainty, which is significantly larger than the systematic uncertainties determined in this paper.  Thus, these data will greatly reduce the uncertainty on the neutrino flux prediction in NuMI due to kaon interactions.

  \section*{Acknowledgments}


We would like to thank the CERN EP, BE and EN Departments for the
strong support of NA61/SHINE.

This work was supported by the Hungarian Scientific Research Fund (Grants
NKFIH 123842--123959), the J\'anos Bolyai Research Scholarship
of the Hungarian Academy of Sciences, the Polish Ministry of Science
and Higher Education (grants 667\slash N-CERN\slash2010\slash0,
NN\,202\,48\,4339 and NN\,202\,23\,1837), the Polish National Center
for Science (grants~2011\slash03\slash N\slash ST2\slash03691,
2013\slash11\slash N\slash ST2\slash03879, 2014\slash13\slash N\slash
ST2\slash02565, 2014\slash14\slash E\slash ST2\slash00018,
2014\slash15\slash B\slash ST2\slash02537 and
2015\slash18\slash M\slash ST2\slash00125, 2015\slash 19\slash N\slash ST2 \slash01689, 2016\slash23\slash B\slash ST2\slash00692),
the Russian Science Foundation, grant 16-12-10176, 
the Russian Academy of Science and the
Russian Foundation for Basic Research (grants 08-02-00018, 09-02-00664
and 12-02-91503-CERN), the Ministry of Science and
Education of the Russian Federation, grant No.\ 3.3380.2017\slash4.6,
 the National Research Nuclear
University MEPhI in the framework of the Russian Academic Excellence
Project (contract No.\ 02.a03.21.0005, 27.08.2013),
the Ministry of Education, Culture, Sports,
Science and Tech\-no\-lo\-gy, Japan, Grant-in-Aid for Sci\-en\-ti\-fic
Research (grants 18071005, 19034011, 19740162, 20740160 and 20039012),
the German Research Foundation (grant GA\,1480/2-2), the
Bulgarian Nuclear Regulatory Agency and the Joint Institute for
Nuclear Research, Dubna (bilateral contract No. 4418-1-15\slash 17),
Bulgarian National Science Fund (grant DN08/11), Ministry of Education
and Science of the Republic of Serbia (grant OI171002), Swiss
Nationalfonds Foundation (grant 200020\-117913/1), ETH Research Grant
TH-01\,07-3 and the U.S.\ Department of Energy.


\bibliographystyle{apsrev4-1}
\bibliography{sample} 

\end{document}